\documentclass[sigconf,nonacm]{acmart}

\AtBeginDocument{%
  }

\usepackage{booktabs}
\usepackage{graphicx}
\usepackage{braket}

\begin{document}

\title{A Framework for Managing the Models of Engineered Quantum Systems}

\author{Siyuan Ji}
\affiliation{%
  \institution{Loughborough University}
  \city{Loughborough}
  \country{UK}
}
\email{s.ji@lboro.ac.uk}

\author{Hayato Ishida}
\affiliation{%
  \institution{Loughborough University}
  \city{Loughborough}
  \country{UK}
}
\email{h.ishida@lboro.ac.uk}

\author{Amal Elsokary}
\affiliation{%
  \institution{Loughborough University}
  \city{Loughborough}
  \country{UK}
}
\email{a.elsokary@lboro.ac.uk}

\author{Stephen Powley}
\affiliation{%
  \institution{Loughborough University}
  \city{Loughborough}
  \country{UK}
}
\email{s.j.powley@lboro.ac.uk}

\author{Catherine White}
\affiliation{%
  \institution{BT Research}
  \city{Ipswich}
  \country{UK}
}
\email{catherine.white@bt.com}

\author{Alexandre Zagoskin}
\affiliation{%
  \institution{Loughborough University}
  \city{Loughborough}
  \country{UK}
}
\email{a.zagoskin@lboro.ac.uk}

\author{Ran Wei}
\affiliation{%
  \institution{Lancaster University}
  \city{Lancaster}
  \country{UK}
}

\email{r.wei5@lancaster.ac.uk}

\author{Maria Aslam}
\affiliation{%
  \institution{University of York}
  \city{York}
  \country{UK}
}
\email{maria.aslam@york.ac.uk}

\author{Michael J. de C. Henshaw}
\affiliation{%
  \institution{Loughborough University}
  \city{Loughborough}
  \country{UK}
}
\email{m.j.d.henshaw@lboro.ac.uk}

\begin{abstract}
Quantum technologies are maturing into systems that classical engineering must build, verify and maintain. The model-driven community has begun to respond with quantum-aware pipelines and languages, and the domain models these produce must be synchronised with the heterogeneous models created and owned by other communities. We argue that existing synchronisation approaches are insufficient for engineered quantum systems. A quantum system description captures superposition and entanglement, which a model transformation could remove undetected, while every structural check passes. To address this, we present the Quantum Systems Model Management (QSysMM) Framework, which guides the construction and synchronisation of the models of a quantum system into a digital single source of truth. The framework features four concerns: ontological, abstraction, composition and exposure, each given the treatment that engineered quantum systems require. Within this framework, we propose a Quantum Systems Modelling Language (QSysML) on the SysML v2 technology stack, and we close with a proposal that matures this synchronisation core into full model management for quantum systems.
\end{abstract}

 

\maketitle
\renewcommand{\shortauthors}{Siyuan Ji et al.}
\section{Introduction}
\label{sec:intro}
\begin{table*}[t]
\caption{Descriptions of a \emph{transmon}, a building block in superconducting quantum computers, from four semantic domains.}
\label{tab:gap}
\small
\begin{tabular}{p{3.2cm}p{10.8cm}}
\toprule
\textbf{Community} & \textbf{Description of a transmon qubit in a bit-flip code} \\
\midrule
Quantum physicist &
Two-level subspace of an anharmonic oscillator;
$H = \hbar\omega_0\, a^\dagger a - \frac{E_C}{2}\, a^\dagger a\,(a^\dagger a - 1)$;
Lindblad operators $L_1 = \sqrt{1/T_1}\,\sigma_-$,
$L_2 = \sqrt{1/2T_\varphi}\,\sigma_z$ with $1/T_\varphi = 1/T_2 - 1/2T_1$;
$\ket{0_L} = \ket{000}$, $\ket{1_L} = \ket{111}$. \\
\addlinespace
Quantum engineer &
$F > 99.5\%$; $T_1 = 50\,\mu s$; $T_2 = 30\,\mu s$; $15\,\text{mK}$; drive at
$5.2\,\text{GHz}$; syndrome via ancilla readout. \\
\addlinespace
Quantum software engineer &
Qubit object in Qiskit; CNOT to ancillas; syndrome decoded by classical
firmware on an FPGA; transpiled to native gates; latency budget $< 2\,\mu s$. \\
\addlinespace
Systems engineer &
Microwave control port (analogue), readout data interface (digital), input line
power $1\,\mu W$; error rate $< 10^{-3}$; verification case VC-03 tests logical
error rate. \\
\bottomrule
\end{tabular}
\end{table*}

Quantum technologies are evolving from laboratory experiments into engineered systems that operate within classical infrastructure. The development of such systems draws on several engineering communities at once. Each community describes the same system in terms of its own semantic domain to represent matters of relevance or importance to its stakeholders (\textbf{concern})s~\cite{iso42010}. To illustrate this, consider transmons, which are superconducting circuits that serve as the physical qubits of the quantum computer, each holding one bit of quantum information. Several transmons are combined to implement a \textbf{bit-flip code}, a simple quantum error-correction scheme that encodes information across several physical qubits so that certain errors can be detected and corrected.  Table~\ref{tab:gap} describes one such transmon from four semantic domains. Although the four descriptions refer to the same physical system, the properties that one community requires are computed in models that another community maintains. The logical error rate that a quantum software engineer reads, for example, depends on physical models outside the software domain. The engineering of the system therefore requires these heterogeneous descriptions to remain consistent with one another.

An established response to this requirement is a single source of truth, one federated model of models from which each community's view can be generated. However, a federated model is only as useful as the synchronisation that maintains the consistency of its constituent models. Without such synchronisation, the federation is representational, a catalogue of descriptions sitting side by side in which a number computed in one model cannot be checked against another. For quantum systems, this synchronisation is where the difficulty lies. The four rows of Table~\ref{tab:gap} are not merely written in different notations; they are four different kinds of mathematical objects, and no mechanism relates them automatically. While individual communities have developed effective techniques for managing the artefacts of their own domains, the engineering of a quantum system requires its models to remain consistent across domain boundaries.

The modelling community has begun to address related issues. Q-READY~\cite{yue2026qready} develops an MBSE pipeline on SysML~v2 that carries a hybrid application from quantum-aware requirements to hardware-aware feasibility, while Gemeinhardt et al.~\cite{gemeinhardt2021} argue for model-driven quantum software engineering; both ask whether a candidate design remains
feasible once physical constraints are considered. Zhao~\cite{zhao2025abstraction} shows that a syntactically valid abstraction can still violate quantum semantics within a quantum software; our work takes that observation to the level of engineered systems models, where transformations cross community boundaries.

More generally, model federation \cite{golra2016,bach2024,amrani2024} integrates heterogeneous models using correspondences (reified links) between elements, while keeping each model in its own space. Model synchronisation \cite{ji2022,batteux2019,stevens2010} then seeks to maintain consistency between such models through transformations that preserve relevant structure. For engineered quantum systems, however, two issues require further treatment. First, a quantum subsystem does not always align with a physical part of the device. A logical qubit, for example, is realised by several physical qubits, and entanglement can link components that a classical model would separate. Second, when a quantum description is abstracted towards a classical description, or composed with one, it may lose superposition, entanglement or correlations, even when ordinary structural checks still pass.

This paper presents QSysMM, a model management framework that organises models of an engineered quantum system around four concerns: ontological, abstraction, composition and exposure. For each concern, the framework identifies the synchronisation that the models require. Relationships between models are represented as typed transformations, each subject to validity conditions grounded in the relevant physics. The aim is to provide a consistent set of models through which the communities building quantum technologies can communicate, and to keep that set synchronised so that each property a stakeholder reads can be traced through checked steps to the physics on which it depends. Examples throughout the paper are drawn from superconducting, gate-based quantum computing. The four concerns, however, are platform-independent. 

\section{The QSysMM Framework}
\label{sec:framework}

We refer to the proposed framework as the \textbf{QSysMM} (Quantum Systems Model Management) Framework. It organises the models of an engineered quantum system as a federated model of models~\cite{bezivin2004}. The framework does not replace the artefacts that communities already maintain; instead, it places them in a shared structure and makes explicit the relationships that must be synchronised.

\begin{figure*}[t]
\centering
\includegraphics[trim=0mm 140mm 90mm 0mm, clip, width=1\textwidth]{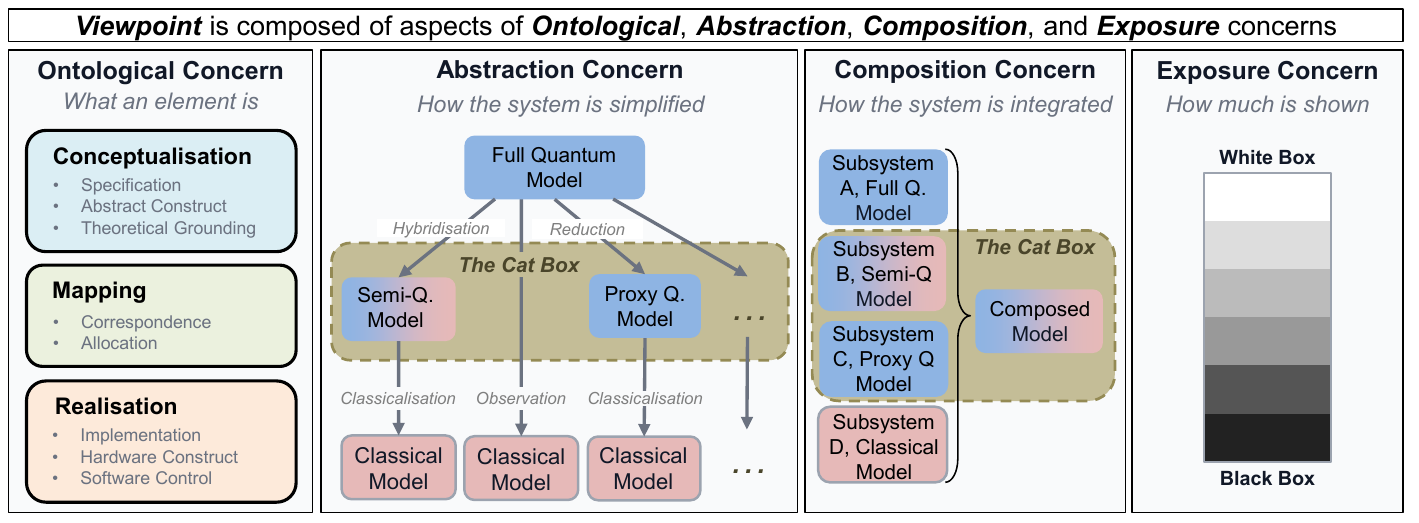}
\caption{The Quantum Systems Model Management (QSysMM) Framework.}
\label{fig:framework}
\end{figure*}

As illustrated in Figure~\ref{fig:framework}, QSysMM is structured by four concerns: ontological, abstraction, composition, and exposure. An \textbf{aspect} is a part of  an entity's  character or nature and can be used for capturing relevant features that help analyse, address or structure one or more \textbf{concern}s.  Here, the \textbf{entities of interest }are models. A \textbf{view} (or architecture view) is a separately identifiable body of information comprising a portion of an \textbf{architecture description}. A \textbf{viewpoint} (or architecture viewpoint) is a set of conventions for the creation, interpretation and use of a view to frame one or more stakeholder concerns~\cite{iso42010}.

The ontological concern determines whether the viewpoint shows Conceptualisation, Realisation, and/or Mapping relations. The abstraction and composition concerns determine which models and transformations are in scope, and exposure determines how much of that material is shown. QSysMM structures aspects of each concern into consistent viewpoints spanning the federated model.

The four concerns adapt familiar model-driven engineering ideas: classifying models, reducing detail, integrating parts and controlling visibility. The following subsections state what each concern and aspect means in QSysMM and where quantum systems require additional checks.

QSysMM represents relationships such as abstraction and composition as typed transformations with contracts. Expressing these contracts is the role of \textbf{QSysML} (Quantum Systems Modelling Language), which we propose on the SysML~v2 technology stack~\cite{omg-sysml2}. QSysML does not reformalise the mathematics, circuit description or simulation that produces a given value; each
remains in its own space, as in Table~\ref{tab:gap}. It carries the transformation and its contract.

\subsection{Ontological concern}
\label{sec:ontological}

The federated model created under this framework is partitioned into three \textbf{ontological aspects}: Conceptualisation, Realisation and the explicit Mapping between them. These aspects do not replace a quantum-computing architecture stack. Existing stacks organise machine operation, for example from
physical qubits through error correction and logical operations to an
application~\cite{jones2012layered}. QSysMM leaves such architectures intact and organises the models that describe them, with Mapping making the relationships between specified behaviour and realised artefacts explicit.

An element belongs to \emph{Conceptualisation} when it states what the system is intended or theoretically expected to do independently of a particular fabricated device. Key elements include the \textbf{specification}, such as a protocol or target logical error rate; the \textbf{abstract constructs} used to state it, such as an ideal quantum algorithm or a logical qubit; and the \textbf{theoretical grounding} that makes the specification achievable, such as a fault-tolerance threshold or an achievable-precision bound. 

An element belongs to \emph{Realisation} when it describes artefacts, controls, measurements or characterised behaviours of the built system. Key elements include the \textbf{implementation}, such as the fabricated device and deployed control firmware; the \textbf{hardware constructs}, such as a transmon with its measured coherence times and its microwave control and readout lines; and the
\textbf{software control} that drives the hardware, such as the pulse sequencer, calibration routines and syndrome decoder.

\emph{Mapping} holds relationships that cross between Conceptualisation and Realisation. A \textbf{correspondence} states that an element in one aspect is the counterpart of an element in the other, such as an ideal logical qubit and the block of physical qubits that implements it. An \textbf{allocation} assigns a specified function to the element that realises it, such as a syndrome-extraction step allocated to an ancilla-based readout and its decoder~\cite{roffe2019}. Unlike a refinement, which proves that a lower-level specification correctly implements a higher-level one~\cite{abadi1991}, mappings defined here cannot be proved. No formal argument links an ideal logical qubit to the transmons that implement it, so the link rests on measurement and must be rechecked as the models change. Mapping is therefore not an extra architectural aspect, but a dedicated model-management layer that builds provenance for the relations that drive the synchronisation between model elements defined in the Conceptualisation and Realisation aspects.

\subsection{Abstraction concern}
\label{sec:abstraction}

\textbf{Abstraction} is a familiar concept in model-driven engineering~\cite{mens2006}, where it is treated as a \emph{vertical} concern that transforms a source model into a reduced model in response to a stakeholder concern. Quantum systems, however, introduce physical properties that might not be preserved by structural abstraction alone. During abstraction, properties such as superposition and entanglement may be discarded without any structural check detecting the loss. To address this challenge, QSysMM defines each abstraction as a transformation equipped with a \textbf{contract}, which contains the provenance that the transformation carries, following the principles of contract-based design~\cite{benveniste2018}. The contract specifies the property preserved, the information discarded, and the validity conditions under which the reduced model remains consistent with its source. This condition is expressed as a constraint over quantities the native model already computes, such as the ratio of two timescales~\cite{reiter2012}. Contract checking, therefore, becomes the model-management task of evaluating that constraint against the values the native model computes. Since those values carry measurement uncertainty, the check is statistical, and a contract must also record a margin as well as a condition. 
We distinguish four representative abstraction operations, grounded in the theory of open quantum systems~\cite{breuer2002}:
\begin{itemize}
    \item \textbf{Reduction:} computes a smaller quantum model by a stated rule, echoing model order reduction in classical systems~\cite{antoulas2005}. As an example, adiabatic elimination~\cite{reiter2012} folds a fast mode into the slow-subspace dynamics, which is valid where the timescales separate.
    \item \textbf{Hybridisation:} treats part of one model classically while the rest stays quantum. As an example, a microwave control field is modelled as a classical drive on a quantum register~\cite{krantz2019}, valid where the drive is strong enough to neglect its quantum fluctuations.
    \item \textbf{Classicalisation:} replaces a decohering model by a classical stochastic one.
    \item \textbf{Observation:} produces a classical outcome by measurment, projective or generalised.
\end{itemize}
These operations form paths through the \emph{cat box}, the part of the model space in which a model carries a recorded loss of quantum content. The full quantum model is outside the box because nothing has yet been discarded. A reduced or hybrid model enters the box when a contract records what was lost and under what condition. Some paths leave the box as classical results, while others remain inside it as still-quantum but already reduced descriptions. Thus, the cat box is not a type of model but a status in a transformation chain, which records that a stakeholder-visible quantity rests on a chain of contracted discards rather than a single calculation. A model can be split into quantum and classical parts in more than one way. These alternatives coexist, each with its own path and contract.

To understand the risk of losing properties through abstraction, consider three physical qubits of the bit-flip code (Table~\ref{tab:gap}), each flipping with probability $q$ per correction cycle. The goal: determine the logical failure rate identified in the Conceptualisation concern. Now consider a reduction that replaces the joint noise model with a product of independent channels, one for each qubit. Every local check still reports the same flip probability $q$, but logical failure depends on the joint behaviour of the qubits. Under independent noise, majority decoding fails only when at least two qubits flip, giving a logical failure rate of approximately $3q^2$. If instead, all three qubits flip collectively, the logical failure rate would be $q$. For $q=10^{-3}$, the failure rates in the two scenarios become $3\times10^{-6}$ and $10^{-3}$, respectively. As a result, verification case VC-03 in Table~\ref{tab:gap}, under the 'Systems Engineers' concern, appears to be satisfied under the reduced model but fails for the source model. Such correlations arise naturally when qubits couple to a shared environment~\cite{mcewen2022} and their impact on fault tolerance is well established~\cite{aharonov2006}. The reduction's contract holds only if the noise acts independently on each qubit, and here it does not. The reduced model, therefore, sits inside the cat box with a violated contract, and the models implemented on the framework can report the failed condition even though every local statistic remains correct.

\subsection{Composition concern}
\label{sec:composition}

Where abstraction follows a vertical chain from a detailed model to a reduced one, \textbf{composition} is a horizontal concern. It integrates distinct subsystem models into a larger description. A real system may have several quantum origins and many classical artefacts, so its full model forms a web of compositions as well as a chain of abstractions.

For classical subsystems, composition can often be handled by model federation~\cite{golra2016,bach2024}: the modeller records how parts correspond and where their interfaces meet. Quantum composition also needs that structural account, but it must add a physical admissibility condition. As with abstraction, QSysMM treats composition as a typed transformation with a contract. The contract captures whether the joint model is the product of its parts or whether it contains content that neither part carries alone. The cat box, therefore, cuts across both the horizontal and vertical axes. A composed model inherits the recorded discards of its inputs, so composing a full quantum model with a model already inside the cat box produces a composed model that also sits inside the cat box. Composition can also create a new cat-box step when the act of joining subsystems assumes away quantum content, for example, by treating entangled subsystems as separable.

The composition depends on whether each subsystem model is still quantum or has already become classical. No special treatment is needed for models that are both classical. Cases requiring explicit treatments are:
\begin{itemize}
    \item \textbf{One quantum, one classical:} a hybrid model that still carries quantum content alongside the classical. The contract states how the classical part conditions or drives the quantum part, for example, a measurement record steering an operation or a control field shaping an evolution.
    \item \textbf{Both quantum (full or reduced):} two models that still carry quantum content compose as a tensor product when they are separable, and when they are entangled, the composed whole carries content that neither part holds.
\end{itemize}

Entanglement has no classical counterpart. Two quantum models may be type-compatible, interface-compatible, and composable under every structural check performed by a modelling tool, yet still compose unsoundly because the systems are entangled while the composition assumes separability. This mirrors the correlated-noise example in Section~\ref{sec:abstraction}, but along the composition axis: separability of the quantum state replaces factorisation of the noise process as the validity condition. Here, QSysMM addresses a fundamentally quantum problem that conventional model management cannot. Existing composition and synchronisation mechanisms reconcile structure and overlap, whereas quantum composition must also satisfy a condition of physical admissibility. Structural compatibility and physical admissibility are independent properties, so the former may hold while the latter does not. The challenge is therefore to express this physical condition within the model, verify it against the native quantum descriptions, and propagate the result throughout the composed model web.

\subsection{Exposure concern}
\label{sec:exposure}

The fourth concern is \textbf{exposure}, which defines how much of a model is shown to a stakeholder, on the scale from black box to white box (Figure~\ref{fig:framework}). What a physicist works with in full may reach an integrator only as a derived value, so exposure lets different communities read the same model at the level of detail they need.

Standard model filtering suffices until the model sits inside the cat box. There, a view may hide not only detail but also a recorded discard and its validity conditions. A black-box view of the logical failure rate in Section~\ref{sec:abstraction} might show $3q^2$ alone; a whiter view would also show that this value rests on a factorised noise model and that the factorisation condition
does not hold. Exposure is therefore tied to contracts for abstraction and composition wherever a model sits inside the cat box, i.e. exposure is how a stakeholder learns whether a value is safe to use. As such, exposure is adjacent to assurance cases, which structure the argument that a claim about a system can be relied upon~\cite{gsn2021}; the checked validity conditions are the evidence such an argument would cite.

\section{Future Plan}
\label{sec:outlook}

This paper has argued that the models of an engineered quantum system cannot be kept consistent by structure-preserving synchronisation and federation alone, because a transformation can discard quantum content while every structural check passes. In response, we have proposed QSysMM, a model management framework that holds each transformation between models to a validity condition that physics decides, drawing on synchronisation, federation and contract-based design for the classical case and adding the physical admissibility that the quantum case requires. We frame the path to a full model management capability as four research questions.

\textbf{RQ1 (Expressibility):} Can the framework be expressed as a reusable library on the SysML~v2 and KerML~\cite{omg-kerml} stack in ways that enable synchronisations with models developed by DSLs from different communities? This question leads to the development of the proposed QSysML as an executable SysML~v2 tooling, using the transmon system of Table~\ref{tab:gap} as a running case study. The implementation includes a fully instantiated view that traces a value property through the four concerns, including alternative abstraction paths.

\textbf{RQ2 (Synchronisability):} Synchronisation between models is central to QSysMM. Once models developed to address the first three concerns (Ontological, Abstraction, and Composition) are integrated, we ask how QSysMM creates synchronisability that extends well beyond a pair of nodes in a model federation. Propagation of changes, therefore, becomes the key test of whether QSysMM scales to complex systems beyond a simple demonstration.

\textbf{RQ3 (Checkability):} Can admissibility conditions be evaluated within model management tooling against quantities computed in native models? Section~\ref{sec:abstraction} argued that a validity condition is a constraint over quantities the native model computes, so the answer is judged on whether evaluating it against those quantities remains practical within an active quantum engineering workflow. The evaluation is also statistical, since witnesses and characterised parameters are estimated from finite data, so the question includes what margins of error a contract must record and how margins compose along a chain. Since deciding separability is computationally hard in general~\cite{gurvits2003}, a composition contract may only be checkable when a witness is available or when the design restricts the class of states under consideration. Characterising this boundary is part of the research question.

\textbf{RQ4 (Extensibility):} Quantum technologies are built on shared physical principles, so it is natural to expect the framework to apply broadly across quantum networking, sensing, and timing systems. This research question asks: Is the QSysMM framework extensible?

\bibliographystyle{ACM-Reference-Format}
\bibliography{refs}

@book{antoulas2005,
  author    = {Athanasios C. Antoulas},
  title     = {Approximation of Large-Scale Dynamical Systems},
  publisher = {SIAM},
  address   = {Philadelphia, PA},
  year      = {2005},
  doi       = {10.1137/1.9780898718713}
}

@inproceedings{bach2024,
  author    = {Jean-Christophe Bach and Antoine Beugnard and Jo\"{e}l Champeau and Fabien Dagnat and Sylvain Gu\'{e}rin and Salvador Mart\'{i}nez},
  title     = {10 Years of Model Federation with Openflexo: Challenges and Lessons Learned},
  booktitle = {Proceedings of the ACM/IEEE 27th International Conference on Model Driven Engineering Languages and Systems (MODELS 2024)},
  publisher = {ACM},
  address   = {New York, NY, USA},
  year      = {2024},
  pages     = {25--36},
  doi       = {10.1145/3640310.3674084}
}

@inproceedings{batteux2019,
  author    = {Michel Batteux and Tatiana Prosvirnova and Antoine Rauzy},
  title     = {Model Synchronization: A Formal Framework for the Management of Heterogeneous Models},
  booktitle = {Model-Based Safety and Assessment (IMBSA 2019)},
  series    = {Lecture Notes in Computer Science},
  volume    = {11842},
  publisher = {Springer},
  address   = {Cham},
  year      = {2019},
  pages     = {157--172},
  doi       = {10.1007/978-3-030-32872-6_11}
}

@article{benveniste2018,
  author    = {Albert Benveniste and Beno\^{i}t Caillaud and Dejan Nickovic and Roberto Passerone and Jean-Baptiste Raclet and Philipp Reinkemeier and Alberto Sangiovanni-Vincentelli and Werner Damm and Thomas A. Henzinger and Kim G. Larsen},
  title     = {Contracts for System Design},
  journal   = {Foundations and Trends in Electronic Design Automation},
  volume    = {12},
  number    = {2--3},
  pages     = {124--400},
  year      = {2018},
  doi       = {10.1561/1000000053}
}

@inproceedings{bezivin2004,
  author    = {Jean B\'{e}zivin and Fr\'{e}d\'{e}ric Jouault and Patrick Valduriez},
  title     = {On the Need for Megamodels},
  booktitle = {Proceedings of the OOPSLA/GPCE Workshop on Best Practices for Model-Driven Software Development},
  address   = {Vancouver, Canada},
  year      = {2004}
}

@book{breuer2002,
  author    = {Heinz-Peter Breuer and Francesco Petruccione},
  title     = {The Theory of Open Quantum Systems},
  publisher = {Oxford University Press},
  address   = {Oxford},
  year      = {2002}
}

@inproceedings{gemeinhardt2021,
  author    = {Felix Gemeinhardt and Antonio Garmendia and Manuel Wimmer},
  title     = {Towards Model-Driven Quantum Software Engineering},
  booktitle = {2021 IEEE/ACM 2nd International Workshop on Quantum Software Engineering (Q-SE)},
  publisher = {IEEE},
  year      = {2021},
  pages     = {13--15}
}

@inproceedings{golra2016,
  author    = {Fahad Rafique Golra and Antoine Beugnard and Fabien Dagnat and Sylvain Gu\'{e}rin and Christophe Guychard},
  title     = {Addressing Modularity for Heterogeneous Multi-model Systems Using Model Federation},
  booktitle = {Companion Proceedings of the 15th International Conference on Modularity (MODULARITY Companion 2016)},
  publisher = {ACM},
  address   = {New York, NY, USA},
  year      = {2016},
  pages     = {206--211}
}

@inproceedings{gurvits2003,
  author    = {Leonid Gurvits},
  title     = {Classical Deterministic Complexity of {Edmonds'} Problem and Quantum Entanglement},
  booktitle = {Proceedings of the 35th Annual ACM Symposium on Theory of Computing (STOC 2003)},
  publisher = {ACM},
  address   = {New York, NY, USA},
  year      = {2003},
  pages     = {10--19},
  doi       = {10.1145/780542.780545}
}

@inproceedings{ji2022,
  author    = {Siyuan Ji and Michael Wilkinson and Charles E. Dickerson},
  title     = {Structure Preserving Transformations for Practical Model-Based Systems Engineering},
  booktitle = {2022 IEEE International Symposium on Systems Engineering (ISSE)},
  publisher = {IEEE},
  year      = {2022},
  pages     = {1--8},
  doi       = {10.1109/ISSE54508.2022.10005437}
}

@article{krantz2019,
  author    = {Philip Krantz and Morten Kjaergaard and Fei Yan and Terry P. Orlando and Simon Gustavsson and William D. Oliver},
  title     = {A Quantum Engineer's Guide to Superconducting Qubits},
  journal   = {Applied Physics Reviews},
  volume    = {6},
  pages     = {021318},
  year      = {2019},
  doi       = {10.1063/1.5089550}
}

@article{jones2012layered,
  author    = {N. Cody Jones and Rodney Van Meter and Austin G. Fowler and Peter L. McMahon and Jungsang Kim and Thaddeus D. Ladd and Yoshihisa Yamamoto},
  title     = {Layered Architecture for Quantum Computing},
  journal   = {Physical Review X},
  volume    = {2},
  pages     = {031007},
  year      = {2012},
  doi       = {10.1103/PhysRevX.2.031007}
}

@article{mens2006,
  author    = {Tom Mens and Pieter Van Gorp},
  title     = {A Taxonomy of Model Transformation},
  journal   = {Electronic Notes in Theoretical Computer Science},
  volume    = {152},
  pages     = {125--142},
  year      = {2006},
  doi       = {10.1016/j.entcs.2005.10.021}
}

@misc{omg-kerml,
  author       = {{Object Management Group}},
  title        = {Kernel Modeling Language ({KerML}), Version 1.0},
  howpublished = {OMG formal specification},
  year         = {2025},
  url          = {https://www.omg.org/spec/KerML/1.0},
  note         = {Formal version published September 2025}
}

@misc{omg-sysml2,
  author       = {{Object Management Group}},
  title        = {{OMG} Systems Modeling Language ({SysML}), Version 2.0},
  howpublished = {OMG formal specification},
  year         = {2025},
  url          = {https://www.omg.org/spec/SysML/2.0},
  note         = {Formal version published September 2025}
}

@article{reiter2012,
  author    = {Florentin Reiter and Anders S. S{\o}rensen},
  title     = {Effective Operator Formalism for Open Quantum Systems},
  journal   = {Physical Review A},
  volume    = {85},
  pages     = {032111},
  year      = {2012},
  doi       = {10.1103/PhysRevA.85.032111}
}

@misc{yue2026qready,
  author        = {Tao Yue and Man Zhang},
  title         = {Q-READY: Predictive Feasibility Assessment for Hybrid Quantum-Classical Applications},
  year          = {2026},
  eprint        = {2606.16201},
  archiveprefix = {arXiv},
  url           = {https://arxiv.org/abs/2606.16201}
}

@article{abadi1991,
  author    = {Mart\'{i}n Abadi and Leslie Lamport},
  title     = {The Existence of Refinement Mappings},
  journal   = {Theoretical Computer Science},
  volume    = {82},
  number    = {2},
  pages     = {253--284},
  year      = {1991},
  doi       = {10.1016/0304-3975(91)90224-P}
}

@article{aharonov2006,
  author    = {Dorit Aharonov and Alexei Kitaev and John Preskill},
  title     = {Fault-Tolerant Quantum Computation with Long-Range Correlated Noise},
  journal   = {Physical Review Letters},
  volume    = {96},
  pages     = {050504},
  year      = {2006},
  doi       = {10.1103/PhysRevLett.96.050504}
}

@article{stevens2010,
  author    = {Perdita Stevens},
  title     = {Bidirectional Model Transformations in {QVT}: Semantic Issues and Open Questions},
  journal   = {Software and Systems Modeling},
  volume    = {9},
  number    = {1},
  pages     = {7--20},
  year      = {2010},
  doi       = {10.1007/s10270-008-0109-9}
}

@techreport{gsn2021,
  author      = {{SCSC Assurance Case Working Group}},
  title       = {Goal Structuring Notation Community Standard, Version 3},
  type        = {{GSN} Community Standard},
  number      = {SCSC-141C},
  institution = {Safety-Critical Systems Club},
  year        = {2021},
  url         = {https://scsc.uk/scsc-141c}
}

@inproceedings{zhao2025abstraction,
  title={When Abstraction Breaks Physics: Rethinking Modular Design in Quantum Software},
  author={Zhao, Jianjun},
  booktitle={2025 40th IEEE/ACM International Conference on Automated Software Engineering (ASE)},
  pages={3886--3890},
  year={2025},
  organization={IEEE}
}

@inproceedings{amrani2024,
  author    = {Moussa Amrani and Rakshit Mittal and Miguel Goul\~{a}o and Vasco Amaral and Sylvain Gu\'{e}rin and Salvador Mart\'{i}nez and Dominique Blouin and Anish Bhobe and Yara Hallak},
  title     = {A Survey of Federative Approaches for Model Management in {MBSE}},
  booktitle = {Proceedings of the ACM/IEEE 27th International Conference on Model Driven Engineering Languages and Systems, Companion (MODELS Companion 2024)},
  publisher = {ACM},
  address   = {New York, NY, USA},
  year      = {2024},
  pages     = {990--999},
  doi       = {10.1145/3652620.3688221}
}

@article{mcewen2022,
  author    = {Matt McEwen and Lara Faoro and Kunal Arya and others},
  title     = {Resolving Catastrophic Error Bursts from Cosmic Rays in Large Arrays of Superconducting Qubits},
  journal   = {Nature Physics},
  volume    = {18},
  pages     = {107--111},
  year      = {2022},
  doi       = {10.1038/s41567-021-01432-8}
}

@misc{iso42010,
  author       = {{International Organization for Standardization}},
  title        = {{ISO/IEC/IEEE} 42010:2022 Software, Systems and Enterprise
                  --- Architecture Description},
  howpublished = {International Standard},
  year         = {2022},
  note         = {ISO/IEC/IEEE 42010:2022}
}

@article{roffe2019,
  author  = {Joschka Roffe},
  title   = {Quantum Error Correction: An Introductory Guide},
  journal = {Contemporary Physics},
  volume  = {60},
  number  = {3},
  pages   = {226--245},
  year    = {2019},
  doi     = {10.1080/00107514.2019.1667078}
}

\end{document}